\documentclass{llncs}

\usepackage{code}
\usepackage{amsmath, amssymb, theorem}
\usepackage{stmaryrd}

\theorembodyfont{\rmfamily\mdseries\upshape}

\newcommand{\mysection}[1]{\vspace*{-4mm}\section{#1}\vspace*{-2mm}}
\newcommand{\mysubsection}[1]{\vspace*{-3mm}\subsection{#1}\vspace*{-1mm}}

        {\hfill $\Diamond$\end{ex}}

\newcommand{\ignore}[1]{}

\renewcommand{\alpha}{a}
\renewcommand{\beta}{b}
\renewcommand{\gamma}{c}
\renewcommand{\tau}{t}

\newcommand{\comment}[1]{}

\newcommand{\LT}{LT}
\newcommand{\simparrow}[0]{\Longleftrightarrow}
\newcommand{\proparrow}[0]{\Longrightarrow}
\renewcommand{\implies}[0]{\supset}
\newcommand{\arrow}[0]{\rightarrow}

\newcommand{\gd}[0]{~\rule{0.5mm}{2.5mm}~}
\newcommand{\tcons}{\, \vdash_{\scriptsize C} \,}
\newcommand{\tdef}{\, \vdash_{\scriptsize R} \,}

\newcommand{\bi}{\begin{array}[t]{@{}l@{}}}
\newcommand{\ei}{\end{array}}
\newcommand{\ba}{\begin{array}}
\newcommand{\ea}{\end{array}}
\newcommand{\bda}{\[\ba}
\newcommand{\eda}{\ea\]}
\newcommand{\bp}{\begin{quote}\tt\begin{tabbing}}
\newcommand{\ep}{\end{tabbing}\end{quote}}

\newcommand{\alphavec}{\bar{\alpha}}

\newcommand{\tenv}{\Gamma}

\newcommand{\tlabel}[1]{\mbox{(#1)}}
\newcommand{\fig}[3]
        {\begin{figure*}[t]#3\
        \caption{\label{#1}#2}\ \hrulefill\ \end{figure*}}

\newcommand{\myirule}[2]{{\renewcommand{\arraystretch}{1.2}\ba{c} #1
                      \\ \hline #2 \ea}}

\newcommand{\llist}[1]{\langle #1 \rangle}

\newcommand{\tv}{\mbox{\it fv}}
\newcommand{\stv}{\mbox{\it \scriptsize fv}}

\newcommand{\turns}{\, \vdash \,}

\newcommand{\sgap}{\quad}
\newcommand{\bgap}{\quad\quad}

\newcommand{\mathem}{\sf}
\newcommand{\IN}{\mbox{\mathem in}}
\newcommand{\LET}{\mbox{\mathem let}}

\newenvironment{ttprog}{\begin{trivlist} %%\small
            \item \tt
        \begin{tabbing}}{\end{tabbing}\end{trivlist}}

\newcommand{\proofsin}[1]{}

\title{Improved Inference for Checking Type Annotations}

\author{Peter J. Stuckey\inst{1}, Martin Sulzmann\inst{2} and Jeremy Wazny\inst{1}}

\institute{
       NICTA Victoria Laboratories \\
       Department of Computer Science and Software Engineering\\
       The University of Melbourne, Vic.\ 3010, Australia\\
       \email{\{pjs,jeremyrw\}@cs.mu.oz.au}
\and
        School of Computing, National University of Singapore \\
        S16 Level 5, 3 Science Drive 2, Singapore 117543 \\
        \email{sulzmann@comp.nus.edu.sg}}

\begin{document}

\maketitle

%%\makeatactive

\bibliographystyle{alpha}

\begin{abstract}
We consider type inference in the Hindley/Milner system 
extended with type annotations and constraints with a particular
focus on Haskell-style type classes.
We observe that standard inference algorithms are incomplete in the presence of nested type annotations.
To improve the situation we introduce a novel inference scheme for checking
type annotations. Our inference scheme is also incomplete in general but
improves over existing implementations as found e.g.~in the
Glasgow Haskell Compiler (GHC). For certain cases (e.g.~Haskell 98) our inference scheme
is complete. Our approach  has been fully implemented 
as part of the Chameleon system (experimental version of Haskell).
\end{abstract}

%------------------------------------------------------------------------------------------------%

\mysection{Introduction}

Type inference for the Hindley/Milner system~\cite{milner:polymorphism} and 
extensions~\cite{Remy!records,pottier-icfp-98,sulzmann-odersky-wehr:journal} of it is a heavily studied area.
Surprisingly, little attention has been given to the impact of type annotations (a.k.a.~user-provided
type declarations) and user-provided constraints on the type inference process.
For concreteness, we assume that the constraint domain is described
in terms of Haskell type classes~\cite{jones:thesis,Hall94}.
Type classes represent a user-programmable constraint domain
which can be used to code up almost arbitrary properties.
Hence, we believe that the content of this paper is of importance for any
Hindley/Milner extension which supports type annotations and constraints.
The surprising observation is that even for ``simple'' type classes type inference
in the presence of type annotations becomes a hard problem.

\begin{example} \label{ex:p}
The following program is a variation of an example from~\cite{haskell-mail}.
Note that we make use of nested type annotations.~\footnote{For concreteness, we annotate {\tt 1} with {\tt Int}
because in Haskell {\tt 1} is in general only a number.}
\bp
class Foo a b where foo ::~a->b->Int\\
instance Foo Int b \\
instance Foo Bool b \\
\\
p y = (\=let \= f ::~c -> Int \\
     \> \>         f x = foo y x \\
   \>    in f, y + (1::Int)) \\
\ep
\bp
q y = (y + (1::Int), \= let \= f ::~c -> Int \\
            \> \>        f x = foo y x \\
           \>     in f) 
\ep
\end{example}
We introduce a two-parameter type class {\tt Foo} which comes with a method {\tt foo}
which has the {\em constrained} type $\forall a,b.Foo~a~b \Rightarrow a\arrow b\arrow Bool$.
The two instance declarations state that $Foo~Int~b$ and $Foo~Bool~b$ hold for any $b$.
Consider functions {\tt p} and {\tt q}.
In each case the subexpression {\tt y+(1::Int)} forces {\tt y} to be of type $Int$.
Note that we could easily provide a more complicated subexpression without type annotations
which forces {\tt y} to be of type $Int$.
The body of function {\tt f x = foo y x} generates the constraint $Foo~Int~t_x$
where $t_x$ is the type of {\tt x}. Note that this constraint is equivalent to $True$ 
due to the instance declaration.
We find that {\tt f} has the inferred type $\forall t_x. t_x \arrow Int$.
We need to verify that this type subsumes the annotated type {\tt f::c->Int}
which is interpreted as $\forall c.c\arrow Int$. 
More formally, we write $C_g \turns \sigma_i \leq \sigma_a$ to denote that the inferred type $\sigma_i$
subsumes the annotated type $\sigma_a$ under some constraint $C_g$. 
Suppose $\sigma_i=(\forall \bar{a}.C_i\Rightarrow t_i)$ and $\sigma_a=(\forall\bar{b}.C_a\Rightarrow t_a)$ 
where there are no name clashes between
$\bar{a}$ and $\bar{b}$. Then, the subsumption condition is (logically) equivalent
to $C_g \models \forall\bar{b}.(C_a \implies (\exists \bar{a}.C_i \wedge t_i=t_a))$. 
In this statement, we assume 
that $\models$ refers to
the model-theoretic entailment relation and $\implies$ refers to Boolean implication.
Outermost universal quantifiers are left implicit.
Note that in our system, we only consider type equality rather than the more general form of subtyping.
For our example, we find that the subsumption condition holds.
Hence, expressions {\tt p} and {\tt q} are well-typed.

Let's see what some common Haskell implementations such as Hugs~\cite{HUGS} and GHC~\cite{GHC} say.
Expression {\tt p} is accepted by Hugs but rejected by GHC whereas GHC accepts {\tt q}
which is rejected by Hugs! Why?

In a traditional type inference scheme~\cite{damas-milner:principal}, constraints are generated while
traversing the abstract syntax tree. At certain nodes (e.g.~let) the
constraint solver is invoked. Additionally, we need to check for correctness of type annotations
(a.k.a.~subsumption check).
The above examples show that different traversals 
of the abstract syntax tree yields different results.
E.g.~Hugs seems to perform a right-first traversal.
We visit {\tt f::c -> Int; f x = foo y x} first without considering the constraints arising out
of the left tuple component. Hence, we find that {\tt f} has the inferred type
$\forall t_x. Foo~t_y~t_x \Rightarrow t_x \arrow Int$ where {\tt y} has type $t_y$.
This type does not subsume the annotated type.
Therefore, type inference fails.
Note that GHC seems to favor a left-first traversal of the abstract syntax tree.

The question is whether there is an inherent problem with nested type annotations,
or whether it's simply a specific problem of the inference algorithms implemented in Hugs and GHC.

\begin{example} \label{ex:no-pt}
Here is a variation of an example mentioned in~\cite{haskell-pt}.
We make use of the class and instance declarations from the previous example.
\bp
\\
test y = \= let \= f ::~c->Int \\
        \> \>    f x = foo y x \\
        \>  in f y 
\ep
\end{example}
Note that {\tt test} may be given types $Int\arrow Int$ and $Bool \arrow Int$.
The constraint $Foo~t_y~t_x$ arising out of the program text can be either
satisfied by $t_y=Int$ or $t_y=Bool$.
However, the ``principal'' type of {\tt test} is of the form
$\forall t_y. (\forall t_x. Foo~t_y~t_x) \Rightarrow t_y \arrow Int$.
Note that the system we are considering does not allow for constraints of the form $\forall t_x.Foo~t_y~t_x$.
%%PJS
Hence, the above example has no (expressible) principal type.
%%PJS For the above example, 
%%PJS we require a universal quantifier in our constraint domain
%%PJS to represent the principal type.
We note that the situation is different for (standard) Hindley/Milner with type annotations.
As shown by Odersky and L{\"a}ufer~\cite{237729}, the problem of finding a solution such that
$\sigma_i$ (the inferred type) is an instance of $\sigma_a$ (the annotated type) can be reduced 
to unification under a mixed prefix~\cite{147067}. 
Hence, we either find a principal solution $\phi$ such that $\turns \phi(\sigma_i) \leq \phi(\sigma_a)$ or no solutions.
Hence, inference for Hindley/Milner with type annotations is complete.

We conclude that type inference for Hindley/Milner with constraints and (nested) type annotations is incomplete.
The incompleteness arises because the subsumption check not only involves a test for correctness of annotations,
but may also need to find a solution. The above example shows that in our general case
there might not necessarily be a principal solution. 

In order to resurrect completeness we could impose a syntactic restriction on the set of programs.
E.g., we could simply rule out type annotations for ``nested'' let-definitions, or
require that the types of all lambda-bound variables occurring in the scope of
a nested annotation must be explicitly provided (although it's unclear whether this is a sufficient condition).
In any case, we consider these as too severe restrictions.

In fact, the simplest solutions seems to be to enrich the language of constraints. Note that the subsumption condition
itself is a solution to the subsumption problem. Effectively, 
we add constraints of the form $\forall\bar{b}.(C_a \implies (\exists \bar{a}.C_i \wedge t_i=t_a))$
to our language of constraints.
Then,  $\forall t_y. (\forall t_x. Foo~t_y~t_x) \Rightarrow t_y \arrow Int$
will become a valid type of {\tt test}
in the above example. This may be a potential solution for some cases, but is definitely undesirable
for Haskell where constraints are attached to dictionaries. It is by no means obvious how to construct
dictionaries~\cite{Hall94} for ``higher-order'' constraints. Furthermore, we believe
that type inference easily becomes undecidable depending on the underlying
primitive constraint domain.

In this paper, we settle for a compromise between full type inference and full type checking.
We only check for the correctness of type annotations. But before checking we infer
as much as possible. Our contributions are:

\begin{itemize}
 \item We introduce a novel formulation of improved inference for checking annotations in terms of
       Constraint Handling Rules (CHRs)~\cite{fruehwirth:CHRs}.
       %%to represent the relations among program text and types. 
      While inferring the type of some inner expression
       we can reach the result of inference for some outer expression.
 \item We can identify a class of programs for which inference is complete.
 \item Our approach is fully implemented as part of the Chameleon system~\cite{chameleon}.
       We can type a much larger class of programs compared to Hugs and GHC.
       E.g., Example~\ref{ex:p} is typable in our system. We refer to~\cite{chameleon} for more examples.
\end{itemize}

We continue in Section~\ref{sec:types-and-constraints} where we formally introduce
an extension of the Hindley/Milner system
with constraints and type annotations. 
Section~\ref{sec:inf-chrs} is the heart of the paper.
We first motivate our approach by example before mapping the entire type inference problem
to CHRs. Type inference then becomes CHR solving.
In Section~\ref{sec:conc} we discuss related work and conclude.

%--------------------------------------------------------------------------------------------------------%

\fig{f:hm-scoped-system}{Hindley/Milner with Type Annotations}{
\bda{c}
    \tlabel{Var-$\forall$E} ~~
    \myirule{(x:\forall\bar{a}. D \Rightarrow t)\in \tenv
        \\ P_p \models C \implies [\bar{t}/\bar{a}]D}
          {C, \tenv \turns x:[\bar{t}/\bar{a}]t}
~~~~
    \tlabel{Abs} ~~
    \myirule{C, \tenv.x:t_1 \turns e:t_2}
          {C,\tenv \turns \lambda x.e: t_1 \arrow t_2}
~~~~
    \tlabel{App} ~~
    \myirule{C, \tenv \turns e_{1}:t_{1} \rightarrow t_{2} \\
           C, \tenv \turns e_{2}:t_{1}}
          {C, \tenv \turns e_{1}~e_{2}: t_{2}}
\\
     \tlabel{Let} ~~
    \myirule{  %%MS: simpler, should still work 
               %%C_2\wedge 
              C_1, \tenv \turns e_1: t_1 \\
            \bar{a} = \tv(C_1,t_1) - \tv(\tenv) \\
           C_2, \tenv.(g:\forall \bar{a}. C_1 \Rightarrow t_1) \turns e_2: t_2}
          {C_2, \tenv \turns \LET\ g=e_1 \, \IN\ e_2: t_2}
~~~~
\tlabel{LetA} ~~
   \myirule{     \bar{a} = \tv(C_1,t_1) 
            \\\ C_2\wedge C_1, \tenv. (g : \forall \bar{a}. C_1 \Rightarrow t_1) 
                        \turns e_1 :  t_1  
           \\ C_2, \tenv. (g : \forall \bar{a}. C_1 \Rightarrow t_1) \turns e_2 : t_2}
           {C_2, \tenv \turns \LET\ 
             \ba{l} g :: C_1 \Rightarrow t_1 \\ g = e_1 \ea ~\IN\ e_2: t_2}
\eda
}

\mysection{Types and Constraints} \label{sec:types-and-constraints}

We present an extension of the Hindley/Milner system
with constraints and type annotations.
\vspace{-4mm}
\bda{llcl}
 \mbox{Expressions} & e & ::= & x \mid \lambda x.e \mid e~e  
                    \mid \LET\ f = e ~\IN\ e \mid 
                 \LET\ \ba{l} f :: C \Rightarrow t  \\ f = e \ea \IN\ e %%%\mid (e:: C \Rightarrow t)
\\ \mbox{Types} & t & ::= & a \mid t \arrow t \mid T~\bar{t}
\\ \mbox{Type Schemes} & \sigma & ::= & t \mid \forall \bar{a}. C \Rightarrow t
\\ \mbox{Constraints} & C & ::= & t=t \mid U~\bar{t} \mid C \wedge C
\\ \mbox{CHRs} & R & ::= & U~\bar{t} \simparrow C \mid 
             U_1~\overline{t_1},..., U_n~\overline{t_n} \proparrow C  
\eda
We write $\bar{o}$ to denote a sequence of objects $o_1,...,o_n$ and
$\overline{o:t}$ to denote $o_1:t_1,...,o_n:t_n$.
W.l.o.g., we assume that 
lambda-bound and let-bound variables have been $\alpha$-renamed to avoid name clashes.
%%MS: redundant
%%We commonly use $x, y, z, \ldots$ to refer to lambda-bound variables and
%%$f, g, h, \ldots$ to refer to user- and pre-defined functions.  Both sets of variables
%%are recorded in a variable environment $\tenv$. 
We record these variables in some environment $\tenv$.
Note that we consider $\tenv$ as an (ordered) list of elements, though we commonly
use set notation. We denote by $\{ x_1 : \sigma_1 ,\ldots, x_n : \sigma_n \} . x:\sigma$
the environment $\{ x_1 : \sigma_1 ,\ldots, x_n : \sigma_n, x:\sigma\}$.

Our type language consists of variables $a$, 
type constructors $T$ and type application, e.g. $T~a$.
We use common notation for writing function and list types.
We also make use of pairs, integers, booleans etc.~in examples.

We find
two kinds of constraints. Equations $t_1=t_2$ among types $t_1$ and $t_2$ and
user-defined constraints $U~\bar{t}$.  We assume that $U$ refers to type classes
such as $Foo$.
For our purposes, we restrict ourselves to {\em single-headed simplification} CHRs
$U~\bar{t} \simparrow C$
and {\em multi-headed propagation} CHRs $U_1~\overline{t_1},..., U_n~\overline{t_n} \proparrow C$.
We note that CHRs describe logic formula.
E.g.~$U~\bar{t} \simparrow C$ can be interpreted as $\forall\bar{a}.U~\bar{t} \leftrightarrow (\exists\bar{b}.C)$
where $\bar{a}=\tv(\bar{t})$ and $\bar{b}=\tv(C)-\bar{a}$, and
$U_1~\overline{t_1},..., U_n~\overline{t_n} \proparrow C$ can be interpreted as
$\forall\bar{a}.(U_1~\overline{t_1}\wedge ...\wedge U_n~\overline{t_n})\implies(\exists\bar{b}.C)$ where
$\bar{a}=\tv(\overline{t_1},...,\overline{t_n})$ and $\bar{b}=\tv(C)-\bar{a}$.
Via CHRs we can model most known type class extensions.
We refer the interested reader to~\cite{overloading-journal}
for a detailed account of translating classes and instances  to CHRs.
We claim that using CHRs we can 
cover a sufficiently large range of 
Hindley/Milner type systems with constraints
%%PJS .
such as functional dependencies, records etc. %%MS: omit now, constructor classes.
Note that CHRs additionally offer 
multi-headed simplification CHRs which in our experience so far
do not seem to be necessary in the type classes context. 
Due to space limitations, we only give one simple example showing how to express
Haskell 98 type class relations in terms of CHRs.

We assume that the meaning of user-defined
constraints (introduced by class and instance declarations)
has already been encoded in terms of some set $P_p$ of CHRs.
User-defined functions are recorded in some initial environment $\tenv_{init}$.
\begin{example}
Consider 
\begin{code}
class Eq a where (==) :: a->a->Bool
instance Eq a => Eq [a]
class Eq q => Ord a where (<) :: a->a->Bool
class Foo a b where foo :: a->b->Int
instance Foo Int b 
instance Foo Bool b
\end{code}
Note that the declaration {\tt class Eq a => Ord a} introduces a new type class $Ord$ and
imposes the additional condition that $Ord~a$ implies $Eq~a$ 
(which is sensible assuming that an ordering relation assumes the existence of an equality
relation). We can model such a condition via a propagation rule.
Hence, $P_p$ consists of
\bda{rrclcrrcl}
 \tlabel{Super} & Ord~a & \proparrow & Eq~a & &  \tlabel{F1} & Foo~Int~b & \simparrow & True \\
 \tlabel{Eq1} & Eq~a & \simparrow & Eq~[a]  &&
 \tlabel{F2} & Foo~Bool~b & \simparrow & True
\eda
and $\tenv_{init} = \{ (==):\forall a.Eq~a \Rightarrow a\arrow a\arrow Bool,
(<):\forall a.Ord~a \Rightarrow a\arrow a\arrow Bool,
foo:\forall a,b.Foo~a~b\Rightarrow a\arrow b\arrow Int \}$.
\end{example}

We introduce judgments of the form $C, \tenv \turns e : t$
where $C$ is a constraint, 
$\tenv$ refers to the set of lambda-bound variables,
predefined and user-defined functions, $e$ is an expression and $t$ is a type. 
We leave the {\em type class theory} $P_p$ implicit.
We say a judgment is {\em valid} iff there is a derivation w.r.t.~the rules found in Figure~\ref{f:hm-scoped-system}.
Commonly, we require that constraints appearing in judgments are satisfiable.
We say that a valid judgment is {\em satisfiable} iff all constraints appearing in the derivation
are satisfiable. A constraint is {\em satisfiable} w.r.t.~a type class theory iff we find some model satisfying
the theory and constraint. We say a theory $P_p$ is {\em satisfiable} iff we find some
model for $P_p$.

In rule \tlabel{Var-$\forall$E}, we assume that $x$ either refers to a lambda- or let-bound variable.
Note that only let-bound variables and primitives can be polymorphic. For convenience, we combine
variable introduction with quantifier elimination.
We can build an instance of a type scheme if the instantiated constraint is entailed by
the given constraint w.r.t.~type class theory $P_p$. 

In rule~\tlabel{Let} we couple the
quantifier introduction rule with 
the introduction of user-defined functions.
In our formulation,
$C_2$ does not necessarily guarantee that $C_1$ is satisfiable.
However, our rule \tlabel{Let} is sound for a lazy semantics which applies to Haskell.

Rule \tlabel{LetA} introduces a type annotation.
Note that we assume that type annotations are {\em closed}, i.e.~all variables appearing
in $C_1 \Rightarrow t_1$ are assumed to be universally quantified.
This is the assumption made for Haskell 98~\cite{haskell98}.
%%In Section~\ref{sec:lexical} we consider an extension where we lift this restriction.
Note that the environment for typing the function body includes
the binding $g : \forall \bar{a}. C_1 \Rightarrow t_1$. Hence, we allow for polymorphic
recursive functions. 

The other rules are standard.
Note that we left out the rule for monomorphic recursive functions
for simplicity.

%-------------------------------------------------------------------------------------------------------%

\mysection{Type Inference via CHRs} \label{sec:inf-chrs}

We introduce our improved inference scheme first by example.
Then, we show how to map the typing problem to a set of CHRs.
We give a description of the semantics of CHRs adapted to our setting.
Finally, we show how to perform type inference in terms of CHR solving.

\mysubsection{Motivating Examples}

The following examples give an overview of the process by which we abstract
the typing problem in terms of constraints and CHRs.

\begin{example}
\label{ex:simple1}
Consider the following program.

\begin{ttprog}
g y = let f x = x in (f True, f y)
\end{ttprog}

We introduce new predicates, (special-purpose) user constraints, $g(t)$ and 
$f(t)$ to constrain $t$ 
to the types of functions {\tt g} and {\tt f} respectively.
It is necessary for us to provide a meaning for these constraints, which
we will do in terms of CHR rules.
The body of each rule will contain all constraints arising from the definition 
of the corresponding function, which represent that function's type.

For the program above we may generate rules similar to the following.
\bda{rcl}
g(t) & \simparrow & t = t_y \arrow (t_1, t_2), f(t_{f1}), 
                    t_{f1} = Bool \arrow t_1, f(t_{f2}),
                    t_{f2} = t_y \arrow f_2 \\
f(t) & \simparrow & t = t_x \arrow t_x
\eda

The arrow separating the rule \emph{head} from the rule \emph{body} can be 
read as logical equivalence. 
Variables mentioned only in a rule's body are implicitly existentially 
quantified.

In the $g$ rule we see that {\tt g}'s type is of the form 
$t_y \arrow (t_1, t_2)$, where $t_1$ and $t_2$ are the results of applying
function {\tt f} to a $Bool$ and a $t_y$.
We represent {\tt f}'s type, at both call sites in the program, by the $f$ user
constraint.

The $f$ rule is much more straightforward. 
It simply states that $t$ is $f$'s type if $t$ is the function type 
$t_x \arrow t_x$, for some $t_x$, which is clear from the definition of 
{\tt f}.

We can infer {\tt g}'s type by performing a CHR \emph{derivation}, solving the
constraint $g(t)$ by applying CHRs (removing the constraint
matching the lhs with the rhs).
Note that we avoid renaming variables where unnecessary.

\bda{rcl}
g(t) & \rightarrowtail_g & t = t_y \arrow (t_1, t_2), f(t_{f1}), 
                           t_{f1} = Bool \arrow t_1, f(t_{f2}),
                           t_{f2} = t_y \arrow f_2 \\
     & \rightarrowtail_f & t = t_y \arrow (t_1, t_2), t_{f1} = t_x \arrow t_x, 
                           t_{f1} = Bool \arrow t_1, f(t_{f2}),
                           t_{f2} = t_y \arrow f_2 \\
     & \rightarrowtail_f & t = t_y \arrow (t_1, t_2), t_{f1} = t_x \arrow t_x, 
                           t_{f1} = Bool \arrow t_1, t_{f2} = t_x' \arrow t_x',
                           t_{f2} = t_y \arrow f_2
\eda

If we solve the resulting constraints for $t$, 
we see that {\tt g}'s type is 
$\forall t_y.t_y \arrow (Bool, t_y)$.

\end{example}

\begin{example}
\label{ex:simple2}
The program below is a slightly modified version of the program presented in
Example~\ref{ex:simple1}.

\begin{ttprog}
g y = let f x = (y,x) in (f True, f y)
\end{ttprog}

The key difference is that {\tt f} now contains a free variable {\tt y}.
Since {\tt y} is monomorphic within the scope of {\tt g} we must ensure that
all uses of {\tt y}, in all definitions, are consistent, i.e. each CHR rule
which makes mention of $t_y$, {\tt y}'s type, must be referring to the same
variable.
This is important since the scope of variables used in a CHR rule is limited
to that rule alone.

In order to enforce this, we perform a transformation akin to lambda-lifting,
but at the type level.
Instead of user constraints of form $f(t)$ we now use binary constraints
$f(t,l)$ where the $l$ parameter represents {\tt f}'s environment.

We would generate rules like the following from this program.
\bda{rcl}
g(t,l) & \simparrow & t = t_y \arrow (t_1, t_2), f(t_{f1}), 
                    t_{f1} = Bool \arrow t_1, f(t_{f2}, \llist{tx}),
                    t_{f2} = t_y \arrow f_2, l = ls \\
f(t,l) & \simparrow & t = t_x \arrow (t_y,t_x), l = (\llist{t_y|ts})
\eda

We write $\llist{t_1,...,t_n}$ to indicate a type-level list containing 
$n$ types.
A list with an $n$-element prefix but an unbounded tail is denoted by 
$\llist{t_1,...,t_n|t}$.
When unifying such a type against another list, $t$ will be bound to some
sublist containing all elements after the $n$th.

As mentioned above, we now use binary predicates to represent the type of a
function.
The first argument, which we commonly refer to as the $t$ component, will
still be bound to the function's type.
The second component, which we call $l$, represents a list of unbound
variables in scope of that function.
We have ensured that whenever the $f$ constraint is invoked 
from the $g$ rule that
$t_y$, the type of {\tt y}, is made available to it.
So, in essence, the $t_y$ that we use in the $f$ rule will have the same type
as the $t_y$ in $g$, rather than simply being a fresh variable known only in
$g$.

\end{example}

\begin{example} \label{ex:p2}
We now return to the program first introduced in Example~\ref{ex:p}, and
generate the CHR rules corresponding to the function {\tt p}, which is
repeated below.
For simplicity we will assume that {\tt (+)} is defined only 
on $Int$s, i.e. $(+) : Int \arrow Int \arrow Int$.

\begin{ttprog}
p y = (\=let \= f ::~c -> Int \\
     \> \>         f x = foo y x \\
   \>    in f, y + (1::Int))
\end{ttprog}

We generate the following CHRs from this fragment of the program.
We also include the rule which represents {\tt foo}'s type, and the rule which
corresponds to the instance $Foo~Int~b$.

\bda{rcl}
p(t,l) & \simparrow & t = t_y \arrow (t_1, t_2), 
                      f(t_1, (\llist{t_y},\llist{t_y,t_x}),
                      t_2 = t_r,  \\
       &            & t_{plus} = Int \arrow Int \arrow Int, 
                      t_{plus} = t_y \arrow Int \arrow t_r,
                      l = (ls, \llist{t_y,t_x}) \\
f_a(t,l) &\simparrow& p(t',l), % _{\ominus},
                      t = c \arrow Int, l = (\llist{t_y|ls},\llist{t_y,t_x}) \\
f(t,l) & \simparrow & t = t_x \arrow t_r, foo(t_{foo},ls'),  
                      t_{foo} = t_y \arrow t_x \arrow t_r, f_a(t,l),  \\
       &            & l = (\llist{t_y|ls}, \llist{t_y,t_x}) \\
foo(t,l)&\simparrow & t = a \arrow b \arrow Int, Foo~a~b \\
Foo~Int~b&\simparrow& True 
\eda

Here we have extended the scheme which we used to generate the constraints in
the previous example. 
We stick to binary predicates, but have expanded the $l$ component to include
two lists.
The first list, which we refer to as the \emph{local} $l$ component contains,
as before, a type-level list of all unbound lambda variables in scope of the
function.
The second list, which we will often denote $LT$ simply contains all of the 
lambda-bound variables from the top-level definition down, in a fixed order.

We introduce a symbol $f_a$ and generate a new rule to represent
{\tt f}'s annotated type.
Note also that we add a call to the $f$ rule to unify {\tt f}'s inferred type 
with the declared type.

As demonstrated earlier, in order to check {\tt f} it is necessary to consider
all of the type information available in {\tt f}'s context. 
In particular, for this program, it is critical that we know {\tt y}
has type {\tt Int}, in order to reduce the $Foo~Int~b$ constraint which arises
from the use of {\tt foo}, but is absent from the annotation.

The way we introduce {\tt f}'s context into {\tt f} is by adding a call from
the $f_a$ rule to the immediate parent definition, which in this case is 
represented by $p$.
In this instance we are not interested in {\tt p}'s type, only the effect it 
has on lambda-bound variables, and any type class constraints which may arise.
Note that if {\tt p} were itself embedded within a function
definition, then it too would have such a call (to its own parent), and so
{\tt f} would indirectly inherit {\tt p}'s context.

We perform the following simplified derivation to demonstrate that the CHR
formulation above captures the necessary context information within {\tt f}.

\bda{rcl}
f(t,l) & \rightarrowtail_f      & foo(t_{foo},ls'), 
                                  t_{foo} = t_y \arrow t_x \arrow t_r,
                                  f_a(t,l), \\
       &                        & l = (\llist{t_y},\llist{t_y,t_x}), ... \\
       & \rightarrowtail_{foo}  & t_{foo} = a \arrow b \arrow Int,
                                  Foo~a~b,
                                  t_{foo} = t_y \arrow t_x \arrow t_r,
                                  f_a(t,l), \\
       &                        & l = (\llist{t_y},\llist{t_y,t_x}), ... \\
                                    
       &                        & \mbox{we can simplify this to:} \\
       &                        & Foo~t_y~t_x,
                                  f_a(t,l), 
                                  l = (\llist{t_y},\llist{t_y,t_x}),
                                  ... \\

       & \rightarrowtail_{f_a}  & Foo~t_y~t_x,
                                  p(t',l), l = (\llist{t_y},\llist{t_y,t_x}),
                                  ... \\
       & \rightarrowtail_{p}    & Foo~t_y~t_x,
                                  t_{plus} = Int \arrow Int \arrow Int,
                                  t_{plus} = t_y \arrow Int \arrow t_r, \\
       &                        & l = (ls, \llist{t_y,t_x}),
                                  l = (\llist{t_y},\llist{t_y,t_x}), ... \\
       & \rightarrowtail_{Foo}  & t_{plus} = Int \arrow Int \arrow Int,
                                  t_{plus} = t_y \arrow Int \arrow t_r, \\
       &                        & l = (ls, \llist{t_y,t_x}),
                                  l = (\llist{t_y},\llist{t_y,t_x}), ... \\
\eda

\end{example}

Through the call to $p$ from $f_a$ we introduce sufficient context information
to determine that $t_y$ is an $Int$, and to consequently reduce away the $Foo$
constraint using the instance rule.
Note that the $LT$ component is necessary here because $p$ is not aware of
its own lambda-bound variables.
Without the $LT$ component, $p$ would not be able to ``export'' the required
information about $t_y$ to $f$.

\mysubsection{Constraint and CHR Generation} \label{sec:formal-inference1}

\fig{f:constraint-lexical2}{Constraint and CHR Generation}{
\mbox{\bf Constraint Generation:} 
\bda{c}
 \tlabel{Var-x} ~~
  \myirule{ (x:t_1) \in \tenv \sgap \mbox{$t_2$ fresh}}
          {\tenv, E, x \tcons (t_2= t_1 \gd t_2)}
~~~~
 \tlabel{Var-f} ~~
  \myirule{\mbox{$t, l, l_g$ fresh} ~~\mbox{$(f \in E$ or} \\  \mbox{$f_a \in E$ $f$ non-recursive)} \\
           C = \{ f(t,l), l = (\llist{\overline{t_x}},l_g)\}}
   { \{ \overline{x:t_x} \} , E, 
  f \tcons ( C \gd t)}
\eda
\bda{c}
 \tlabel{VarA-f} ~~
  \myirule{\mbox{$f_a\in E$  $f$ recursive} \sgap \mbox{$t, l, l_g$ fresh} \sgap
           C = \{ f_a(t,l), l = (\llist{\overline{t_x},l_g})\}}
   {  \{ \overline{x:t_x} \} , E, 
  f \tcons ( C \gd t)}

\vspace{1mm}
\\ 
\tlabel{Abs} ~~
  \myirule{%%\tv(t_1) \subseteq V \\
            \tenv.x:t_1, E,e \tcons (C \gd t_2) \\ C'=\{C, t_3=t_1\arrow t_2 \} \sgap \mbox{$t_3$ fresh}}
          {\tenv, E,\lambda x::t_1.e \tcons (C' \gd t_3)}
~~~~
\tlabel{App} ~~ \myirule{\tenv, E,e_1 \tcons (C_1 \gd t_1) 
          \\ \tenv, E,e_2 \tcons (C_2 \gd t_2)
       \sgap \mbox{$t_3$ fresh}}
         {\tenv, E,e_1~e_2 \tcons 
       (C_1,C_2,t_1=t_2 \arrow t_3 \gd t_3)}

\vspace{1mm}
\\
\tlabel{Let} ~~ \myirule{\tenv, E\cup\{g\},e_2 \tcons (C \gd t)}
         {\tenv, E, \LET\ g = e_1 ~\IN\ e_2 \tcons (C \gd t)}
~~~~
\tlabel{LetA}  ~~
\myirule{        \tenv, E\cup \{g_a\}, e_2 \tcons (C \gd t)}
             {\tenv,   E,\LET\ 
             \ba{l} g :: C_1 \Rightarrow t_1 \\ g = e_1 \ea ~\IN\ e_2 
           \tcons (C \gd t)}
\eda
\mbox{\bf CHR Generation:} 
\bda{c}
  \tlabel{Var} ~~
  h, \tenv, E,v \tdef \emptyset
~~~~
 \tlabel{App} ~~
  \myirule{h, \tenv, E,e_1 \tdef P_1 \\ h, \tenv, E,e_2 \tdef P_2}
          {h, \tenv, E,e_1~e_2 \tdef P_1 \cup P_2}
~~~~
 \tlabel{Abs} ~~
 \myirule{h, \tenv.x:t, E,e \tdef P }
         {h, \tenv, E,\lambda x::t.e \tdef P}

\vspace{2mm}
\\ 
 \tlabel{Let} ~~
   \myirule{      \tenv = \{ x_1 : t_1,\ldots,x_n : t_n \}
           \sgap \mbox{$t,t_2',l,lr,lg$ fresh} \\ g,\tenv, E, e_1 \tdef P_1 
             \sgap h, \tenv, E\cup\{g\}, e_2 \tdef P_2 \sgap  \tenv, E,e_1 \tcons (C_1' \gd t_1') 
          \\ P = P_1 \cup P_2 \cup 
        \{ g(t,l) \simparrow C_1',t_1'=t,l=(\llist{t_1,\ldots,t_n | lr},\LT), h(t_2',l)_{\ominus} \} }
         {h, \tenv,E,\LET\ g = e_1 ~\IN\ e_2 \tdef P}

\vspace{2mm}
 \\ 
 \tlabel{LetA} ~~~
  \myirule{ 
            \tenv = \{ x_1 : t_1,\ldots,x_n : t_n \}            \sgap \mbox{$t,t_2',l,l_r$ fresh}
         \\ g, \tenv, E\cup\{g_a\}, e_1 \tdef P_1 \sgap h, \tenv, E\cup\{g_a\}, e_2 \tdef P_2
          \sgap \tenv, E\cup\{g_a\}, e_1 \tcons (C'_1 \gd t'_1)
          \\    P= P_1 \cup P_2 \cup \\
         \left \{ \ba{lcl}
       g_a(t,l) & \simparrow  & t=t''_1, C''_1, l=(\llist{t_1,...,t_n|lr},\LT), h(t_2',l)_{\ominus}
       \\ g(t,l) & \simparrow &   
       l=(\llist{t_1,\ldots.t_n | lr},\LT ), g_a(t,l), C'_1, t=t'_1 
           \ea
          \right \} 
       }
        {h, \tenv,   E,\LET\ 
             \ba{l} g :: C''_1 \Rightarrow t''_1 \\ g = e_1 \ea ~\IN\ e_2 \tdef P}
\eda
}
%
%%\ms{BTW, in rule \tlabel{Var-Annot}, we could argue if we'd use rule ($f$) instead, we sometimes might get better
%%error messages. Ok, we need to be careful not to get into a cycle.}

In detail, we show how to map expressions to constraints and CHRs.
Lambda-abstractions such as $\lambda x.e$ are preprocessed and turned into
$\lambda x::t_x.e$ where $t_x$ is a fresh type variable.
We assume that $\LT$ contains all such type variables $t_x$ attached to lambda-abstractions.

For each function definition {\tt f=e} we generate a CHR of the form
$f(t,l) \simparrow C$ where $l$ refers to a pair $(l_l,l_g)$.
%%Silently, we assume that predicate symbols referring to function types and annotations
%%are now binary.
The constraint $C$ is generated out
of the program text of {\tt e}.
We maintain that $l_l$ denotes the 
set of types of lambda-bound variables in the environment and $l_g$ refers
to $\LT$ the types of all lambda-bound variables.

We make use of list notation (on the level of types) to refer to
the types of $\lambda$-bound variables. In order to avoid confusion
with lists of values, we write $\llist{l_1,\ldots,l_n}$ to denote the list of
types $l_1,\ldots,l_n$.  We write $\llist{l | r}$ 
to denote the list of types with head $l$ and tail $r$.

For constraint generation, we employ judgments $\tenv, E, e \tcons (C \gd t)$
where environment $\tenv$, set of predicate symbols $E$ and expression $e$ are input values and
constraint $C$ and type $t$  are output parameters.
Note that $\tenv$ consists of lambda-bound variables only
whereas $E$ holds the set of predicate symbols referring to
primitive and let-defined functions.
Initially, we assume that $E_{init}$ holds all the symbols defined in $P_{init}$ which is the
CHR representation of all functions in $\tenv_{init}$.
The rules can be found in Figure~\ref{f:constraint-lexical2}.

Consider rule \tlabel{Var-f}.
If the function does not carry an annotation or the function is not recursive~\footnote{We can easily check whether a function
is recursive or not by a simple dependency analysis.} we make use of the definition CHR.
%%Note that the (logical reading of the) annotation CHR implies the definition CHR.
However, strictly making use of the definition CHR might introduce cycles among CHRs, e.g.~consider
polymorphic recursive functions. In such cases we make use of the annotation CHR, see rule~\tlabel{VarA-f}. 
%%MS: omit, but elaborate in this point in journal version
%%Note that in the logical inference system (see Figure~\ref{f:inf-logic-formula}) we strictly make use
%%of the predicate describing the annotation. 
%%However, we obtain a slightly more expressive CHR inference scheme if we make
%%use of the definition CHR in case of a type annotated non-recursive function (see upcoming Example~\ref{ex:ex}).
%%MS: note that in the logical inference system we can easily reach surrounding information
In both rules we set $l$ to the sequence of types of all lambda-variables in scope.
Note that we might pass in more types of lambda-bound variables
than expected by that function. This is safe because we leave the first component of $l$ ``open''
at definition sites. That is, we expect at least the types of lambda-bound variables in
scope at the definition site and possibly some more. The second component which refers to the
sequence of all types of lambda-bound variables appearing in the entire program is left unconstrained.
This component will be only constrained at definition sites (see CHR generation rules~\tlabel{Let} and~\tlabel{LetA}).
Note that in rule \tlabel{Abs} the order of lambda-bound variables added to type environment matters.
Hence, we silently treat $\tenv$ as a list rather than a set.
In rules \tlabel{Let} and \tlabel{LetA} the constraints arising out of $e_1$ might not appear
in $C$ unless we use function $g$ in $e_2$.
Note that we do not generate a constraint for the subsumption condition which will be checked separately.

For rule generation, we employ judgments of the form $h,\tenv, E, e \tdef P$
where CHR $h$, environment $\tenv$, set of predicate symbols $E$
and expression $e$ are input values and the set $P$ of CHRs is the output value.
As an invariant we maintain that $h$ refers
to the surrounding definition of expression $e$.
Initially, we assume that $h$ refers to some 
trivial CHR $h(t,l)\simparrow True$ and $E$ refers to the set of primitive functions.
We refer to Figure~\ref{f:constraint-lexical2} for details. There are two interesting rules.

Rule \tlabel{Let} deals with unannotated functions.
Note that we do not add $g$ to $E$ when generating constraints and rules from $e_1$.
Hence, we assume for simplicity that unannotated functions are not allowed to be recursive. 
Of course, our system~\cite{chameleon} handles unannotated, recursive functions.
Their treatment is described in a forthcoming report
%%MS: todo, add ref to report
%%We address such cases in Appendix~\ref{sec:mon-rec-func}.
The novel idea of our inference scheme is
that we reach surrounding constraints within the definition of {\tt g}
via the constraint $h(t_2,l)_{\ominus}$. 
The $_{\ominus}$ marker (left out in Example~\ref{ex:p2} for simplicity) 
serves two purposes: (1) We potentially create cycles among CHRs because we might reintroduce renamed
copies of surrounding constraints. 
Markers will allow us to detect such cycles to avoid non-termination of CHRs.
(2) Variables occurring in marked constraints are potentially part of the environment.
Hence, we should not quantify over those variables.
Examples will follow shortly to highlight these points.

%%MS: safe space
%%Note that the first component of
%%$l$ refers to the sequence of types of lambda-bound variables and scope and possibly some more
%%whereas the second component refers to $\LT$, the set of types of all lambda-bound variables.

Rule \tlabel{LetA} is similar to rule \tlabel{Let}.
Here, the annotation CHR includes the surrounding definition $h$. 
The actual inference result is reported in the definition CHR.

%%MS: moved this discussion to Appendix
%%As mentioned in the Introduction, including the surrounding information on the rhs of
%%{\tt g}'s definition CHR yields a weaker inference system (see upcoming Example~\ref{ex:spurious}).

\mysubsection{CHR Solving} \label{sec:formal-inference2}

We introduce the marked CHR semantics.
We assume that each constraint is attached with either a $_{\ominus}$ marker 
or a $_{\epsilon}$ (pronounced empty) marker.
The empty marker is commonly left implicit. We refer to constraints carrying a $_{\ominus}$ marker as {\em marked} constraints.
A constraint carrying the empty marker is {\em unmarked}.
In case of CHR rule application on a marked constraint, we mark all constraints 
in the body before adding them to the constraint store.

\begin{definition}[Marked CHR Application]
Let  $d= (U~\bar{t})_a$ (or $d= f(\bar{t})_a$) be a primitive constraint where $a\in \{ _{\ominus}, _{\epsilon} \}$. 
We define $mark(d)=a$. 
We write $d_{\ominus}$  to denote $(U~\bar{t})_{\ominus}$ (or $f(\bar{t})_{\ominus}$).

Let $(R)~~c_1,...,c_n \proparrow d_1, ..., d_m \in P$ and $C$ be a constraint.
                   Let $\phi$ be the m.g.u.~of all equations in $C$. Let $c'_1,...,c'_n\in C$ such that there exists a substitution $\theta$
                   on variables in rule (R) such that $\theta(c_i)=\phi(c'_i)$ for $i=1...n$. 
                    Then, $C \rightarrowtail_R C,d'_1,...,d'_m$
                   where \bda{lcl}  d'_i & = & 
                             \left \{ \ba{ll} d_i & \mbox{if $mark(c'_j) \not= {}_{\ominus}$ for $j=1...n$} \\
                                              (d_i)_{\ominus} & \mbox{otherwise}
                                      \ea
                             \right .
                         \eda

Let $(R)~~c \simparrow d_1, ..., d_m \in P$ and $C$ be a constraint.
                   Let $\phi$ be the m.g.u.~of all equations in $C$. Let $c'\in C$ such that there exists a substitution $\theta$
                   on variables in rule (R) such that $\theta(c)=\phi(c')$, that is user-defined constraint $c'$ {\em matches}
                   the left-hand side of rule (R). Then, $C \rightarrowtail_R C-c',d'_1,...,d'_m$
                   where $d'_i$ are as above.
\end{definition}
A derivation step from global set of constraints $C$ to $C'$
using an instance of rule $r$ is denoted $C \rightarrowtail_r C'$.
A \emph{derivation}, denoted $C \rightarrowtail_P^* C'$
is a sequence of derivation steps using either rules in $P$ such that
no further derivation step is applicable to $C'$.
The operational semantics of CHRs exhaustively apply rules to the
global set of constraints, being careful
not to apply propagation rules twice on the same constraints (to avoid
infinite propagation).  
%%MS: let's save space
%%For more details on avoiding 
%%repropagation see e.g.~\cite{abdennadher:confluence}.
We say a set of CHRs is {\em terminating} if for each $C$ there exists $C'$ such 
that $C \rightarrowtail^*_P C'$.
%% MS: redundant
%%We say a constraint $C$ is {\em satisfiable} if there exists $C'$ such that $C \rightarrowtail^*_P C'$
%%and the unifier of all equations in $C'$ exists.

\begin{example} \label{ex:trouble-sub-check}
Consider
\begin{ttprog}
class Erk a where erk ::~a \\
class Foo a where foo ::~a \\
f = (erk, let g ::~Foo a => a; g = foo in g) 
\end{ttprog}
\end{example}
Here is a sketch of the translation to CHRs.
\bda{rclcrcl}
 g_a(t) & \simparrow & f(t')_{\ominus}, Foo~t &$\sgap$ &  f(t)   & \simparrow & t=(a,b), Erk~a, g(b)\\
 g(t)   & \simparrow & g_a(t), Foo~t \\
\eda

Consider the derivation
\bda{ll}
 & \underline{g(t_1)} \rightarrowtail_g   ~g_a(t_1), Foo~t_1 
~\rightarrowtail_{g_a} ~ \underline{f(t')_{\ominus}}, Foo~t_1  \\
\rightarrowtail_{f} &
            t'=(a,b), (Erk~a)_{\ominus}, g(b)_{\ominus}, Foo~t_1 \\
\rightarrowtail_g & t'=(a,b), (Erk~a)_{\ominus}, \underline{g_a(b)_{\ominus}}, (Foo~b)_{\ominus}, Foo~t_1 
\eda

In step $\rightarrowtail_f$ we propagate $\ominus$ to all new constraints.
%we overwrite the rhs with $_{\ominus}$.
Note that we encounter a cycle among CHRs (see underlined constraints). 
Indeed, CHRs may be ``non-terminating'' because we introduce repeated duplicates
of surrounding constraints.
%%Note that $t'$ and $t''$ are fresh variables which are not constrained by any of the variables in the initial constraint store.
To avoid non-termination we introduce an additional
{\em C}HR {\em C}ycle {\em R}emoval step
\bda{ll}
 ... \\
\rightarrowtail_g & t'=(a,b), (Erk~a)_{\ominus}, \underline{g_a(b)_{\ominus}}, (Foo~b)_{\ominus}, Foo~t_1  \\
\rightarrowtail_{CCR} & t'=(a,b), (Erk~a)_{\ominus}, (Foo~b)_{\ominus}, Foo~t_1
\eda
which is defined as follows.

\begin{definition}[CHR Cycle Removal] \label{def:chr-cycle-removal}
Let $f(t,l) \in C$ and $f(t',l')_{\ominus} \in C'$ and a derivation
            $... \rightarrowtail C \rightarrowtail ... \rightarrowtail C'$.
  Then, $... \rightarrowtail C \rightarrowtail ... \rightarrowtail C' \rightarrowtail_{CCR} C'-f(t',l')_{\ominus}$.

We assume that $\rightarrowtail_{CCR}$ is applied aggressively.
\end{definition}

We argue that this derivation step is sound because any further rule application on $g_a(b)_{\ominus}$ will only add
renamed copies of constraints already present in the store.

\begin{lemma}[CCR Soundness] \label{le:rmc-soundness}
Let $P$ be a set of CHRs and $C$ and $C'$ two constraints such that $C\rightarrowtail^*_P C'$.
Then $P \models C \leftrightarrow \bar{\exists}_{\stv(C)} C'$.
\end{lemma}
%%MS:
%% consider cycle among two marked function predicates $f(t,l)_{\ominus}$ and $f(t',l')_{\ominus}$
%% constraints connected to t' are renamed copies of constraints connected to t
%% same applies to constraints affecting lambda-bound variables

We can also argue that we break any potential cycle among predicate symbols referring to function symbols.
Note that we do not consider breaking cycles among two unmarked constraints. Such cases will only occur
in case of unannotated, recursive functions which are left out for simplicity.
A detailed description of such cases will appear in a forthcoming report.
%%MS: todo, discuss in report in detail
%%We discuss removal of such cycles in Appendix~\ref{sec:mon-rec-func}.
Also note that we never remove cycles in case of user-defined constraints. 
In such a case, the type class theory
might be non-terminating.
Hence, we state the the CCR derivation steps preserves termination assuming the type class theory is terminating.

\begin{lemma}[CCR Termination] \label{le:cchr-termination}
Let $P_p$ be a terminating type class theory, $h(t,l)\simparrow True$ a CHR, $E_{init}$ a set of
primitive predicate symbols, $P_{init}$ a set of CHRs  and
$(\tenv,e)$ a typing problem
such that $h, \tenv, E_{init}, e \tdef P_e$ and $(\tenv_{init},P_{init})$ models $E_{init}$
for some $\tenv_{init}$. Then, $P_p\cup P_e \cup P_{init}$ is terminating.
\end{lemma}

\mysubsection{Type Inference via CHR Solving} \label{sec:formal-inference3}

Consider type inference for an expression $e$ w.r.t.~an environment $\tenv$ of lambda-bound variables
and an environment $\tenv_{init}$ of primitive functions and
type class theory $P_p$. We assume $(P_{int},E_{int})$ model $\tenv_{init}$
such that for each $f:\forall \bar{a}.C \Rightarrow t'$ we find $f(t,l) \simparrow C,t=t' \in P_{init}$
and $f\in E_{init}$.
Then, we generate $\tenv, e \tcons (C \gd t)$ and 
$h,\tenv, E_{init}, e \tdef P_e$. We generally assume that $P$ denotes $P_p \cup P_e \cup P_{init}$.

For typability we need to check that (1) constraint $C$ is satisfiable, and
(2) all type annotations in $e$ are correct.
We are now in the position to describe CHR-based satisfiability and subsumption
check procedures.

\begin{definition}[Satisfiability Check] \label{def:sat} \mbox{} \\
Let $P$ be a set of CHRs and $C$ a constraint such that
$C \rightarrowtail^*_P C'$ for some constraint $C'$.
We say that $C$ is satisfiable iff the unifier of all equations in $C'$ exists.
\end{definition}
Soundness of the above definition follows from results stated in~\cite{overloading} in combination
with Lemma~\ref{le:rmc-soundness}.
%%MS: see proof of Le1
Of course, decidability of the satisfiability check depends on whether CHRs are terminating.

To check for correctness of type annotations we first need to calculate the set
of all subsumption problems. 
Let $E_{sub(e)}$ be the set of all predicate symbols $g_a$ where each $g_a$ refers to
some subexpression $(\LET\ g :: C_1 \Rightarrow t_1; g = e_1 ~\IN\ e_2)$ in $e$.
Let $F_{sub(e)}$ be a formula such that $\forall t,l.(g_a(t,l)\leftrightarrow g(t,l)) \in F_{sub(e)}$
for all $g_a \in E_{sub(e)}$.
It remains to verify that  the type annotation is correct under the abstraction of type inference in terms of $P$.
Formally, we need to verify that
$P \models F_{sub(e)}$ where $P$ refers to first-order logic 
interpretation of the set of CHRs $P$. 
In~\cite{overloading}, we introduced a CNF (Canonical Normal Form) 
procedure to test for equivalence among constraints $\forall t,l.(g_a(t,l)\leftrightarrow g(t,l))$ w.r.t.~some
set of CHR by executing $g_a(t,l)$ and $g(t,l)$ and verify that the resulting
final stores are equivalent modulo variables in the initial store (here $\{t,l\}$).
Thus, we can phrase the subsumption check as follows. We write $\bar{\exists}_{t,l}.C$ to denote
$\exists\tv(C)-\{t,l\}.C$.
%%MS: use now <-> from the start
%%Note that $(\forall t,l.g_a(t,l)\implies g(t,l))$ iff 
%%$(\forall t,l.g_a(t,l)\leftrightarrow g(t,l)$ w.r.t.~$P$. 
%%Recall that we include the annotation CHR on the rhs of each definition CHR.

\begin{definition}[Subsumption Check] \label{def:sub}
Let $g_a\in E_{sub(e)}$ and $P$ be a set of CHRs.
We say that {\tt g}'s annotation is correct iff
(1) we execute $g_a(t,l) \rightarrowtail^*_P C_1$
and $g(t,l) \rightarrowtail^*_P C_2$,
(2) we have that $\models (\bar{\exists}_{t,l}.C_1) \leftrightarrow  (\bar{\exists}_{t,l}.C_2)$.
\end{definition}
Soundness of the above definition follows 
from results stated in~\cite{overloading} in combination
with Lemma~\ref{le:rmc-soundness}.
%%MS: see proof of Le4

\begin{example} \label{ex:erk-foo-check}
Recall Example~\ref{ex:trouble-sub-check} and the derivation
$g(t_1) \rightarrowtail^* t'=(a,b), (Erk~a)_{\ominus}, Foo~t_1$.
A similar calculation shows that $g_a(t_1) \rightarrowtail^* 
s'=(c,d), (Erk~c)_{\ominus}, Foo~t_1$.
Note that the resulting constraints
are logically equivalent modulo variable renamings.
Hence, {\tt g}'s annotation is correct.
\end{example}

Note that in our formulation of type inference, the type of an expression is described by a set of
constraints w.r.t.~a set of CHRs. The following procedure describes how to build
the associated type scheme.
Markers attached to constraints provide important information which variables arise from the surrounding
scope. Of course, we need to be careful not to quantify over those variables.

\begin{definition}[Building of Type Schemes]
Let $P$ be a set of CHRs, {\tt g} a function symbol.
We say function {\tt g} has type $\forall\alphavec.C'\Rightarrow t'$ w.r.t.~$P$ iff
(1) We have that $g(t, l) \rightarrowtail^* C,l=(l_l,l_g)$ for some constraint $C$,
(2) $\phi$, the m.g.u.~of $C,l=(l_l,l_g)$ exists,  
(3) let $D\subseteq \phi(C)$ such that
$D$ is maximal and $D$ consists of unmarked user-defined constraints only,
(4) let $\bar{a} = \tv(D, \phi t) - \tv(\phi l_l)$,
(5) let $C'=\phi(C)$ and $t'=\phi(t)$.
\end{definition}

\begin{example}
According to Example~\ref{ex:erk-foo-check}
we find that 
{\tt g} has type $\forall t_1.(Erk~a,Foo~t_1)\Rightarrow t_1$.
Note that $Erk~a$ arises from {\tt f}'s program text.
\end{example}

\comment{
%%MS: omit, unnecessary
The informed reader might know that
in~\cite{overloading} we defined subsumption checking in terms of type schemes. 
This check is no longer appropriate for subsumption checking
in case of nested definitions.
Applying the approach in~\cite{overloading} would result in checking
whether $\forall t_1.(Erk~a, Foo~t_1)\Rightarrow t_1$
subsumes $\forall t_1.(Erk~c, Foo~t_1)\Rightarrow t_1$.
This is obviously not the case unless we rename the global constraints (e.g., take $c=a$).
However, it is not obvious how to find the renaming substitution.
Note that our new subsumption check (see Definition~\ref{def:sub})
is free of such problems.
}

We are able to state soundness of our approach.

\begin{theorem}[Soundness]
Let $P_e$, $P_p$ and $P_{init}$ be three sets of CHRs, $h$ a CHR in $P_{init}$, 
$\tenv$ an environment of simply-typed bindings,
$\tenv_{init}$ an environment of primitive functions, 
$e$ an expression,
%%MS: not necessary
%%$e$ a let-realizable expression, 
$E_{init}$ a set of predicate symbols,
$C$ a constraint and $t$ a type
such that $(P_{init}, E_{init})$ models $\tenv_{init}$ and
$\tenv, E_{init}, e \tcons (C \gd t)$ and $h,\tenv, E_{init}, e \tdef P_e$
and type checking of all annotations in $e$ is successful.
Let $C \rightarrowtail^*_{P_p \cup P_{init} \cup P_e} C'$ for some constraint $C'$.
Let $\phi$ be the m.g.u.~of $C'$ where we treat all variables in $\tenv$ as Skolem constants.
Then, $\phi(C'), \phi(\tenv)\cup\tenv_{init} \turns e : \phi(t)$.
\end{theorem}
\comment{
\begin{proof}
CHRs entail logic formula, hence, we immediately obtain soundness.
\end{proof}
}

The challenge is to identify some sufficient criteria under which our type inference 
method is complete. Because we only {\em check} for subsumption we need to
guarantee that each subsumption condition will be either true or false.
E.g. in Example~\ref{ex:no-pt} the subsumption condition boils down to the constraint
$\forall t_x.Foo~t_y~t_x$. Note that we can satisfy this constraint by setting $t_y$ to
either $Int$ or $Bool$. Hence, our task is to prevent such situations from happening.
In fact, such situations can never happen for single-parameter type classes.
But what about multi-parameter type classes?
The important point is to ensure that fixing one parameter will immediately fix all the others.
That is, in case of $\forall t_x.Foo~t_y~t_x$ we know that $t_x$ is uniquely determined by $t_y$.
We can enforce such conditions in terms of functional dependencies~\cite{JonesESOP2000}.

\begin{definition}
We say a type class $TC$ is {\em fully functional} iff
we find a class declaration
{\tt class TC a$_1$ ... a$_n$ | fd$_1$,..., fd$_n$}
where {\tt fd$_i$} = {\tt a$_i$ -> a$_1$ ... a$_{i-1}$ a$_{i+1}$ ... a$_{n}$}. 
\end{definition}

We argue that for fully functional dependencies solutions (if they exist) must be unique.
In fact, this is not sufficient because there are still cases where we need guess.

\begin{example} \label{ex:ambig}
Consider the following simplified representation of the Show class.
\begin{code}
class Show a where
  show :: a->String
  read :: String->a
f :: Show a => String->String
f x = show (read x)
\end{code}
The subsumption check boils down to the formula
$\forall a.Show~a \implies \exists a'.Show~a'$ which is obviously a true statement (take $a'=a$).
However, in our translation to CHRs we effectively check for $Show~a \leftrightarrow Show~a'$
which obviously does not hold.
\end{example}

There are further sources where we need to take a guess.
\begin{example} \label{ex:incomplete}
Consider
\begin{code}
class Foo
instance Show Int
instance Foo a => Show a
\end{code}
where $P_p = \{ \tlabel{S1} ~Show~Int \simparrow True \gd \tlabel{S2}~ Show~a \simparrow Foo~a \}$.
We have that $P_p\models Show~Int$. 
However, $Show~Int \rightarrowtail_{S2} Foo~Int$ where $\models Foo~Int \not\leftrightarrow True$
which suggests that $P_p \models Show~Int$ might not hold. Clearly, by guessing the right path
in the derivation we find that $Show~Int \rightarrowtail_{S1} True$.
\end{example}

To ensure that our subsumption check (Definition~\ref{def:sub}) is complete
we need to rule out ambiguous types and require that the type class theory is complete.
A type is {\em ambiguous} iff
we can not determine the variables appearing in constraints by looking at the types alone.
The annotation {\tt f::Show a=>String->String} in Example~\ref{ex:ambig} is ambiguous.
A type class theory $P_p$ is {\em complete} iff $P_p$ is 
confluent, terminating and range-restricted (i.e.~grounding the lhs of CHRs grounds the rhs)
and all simplification rules are single-headed. The type class theory $P_p$ in Example~\ref{ex:incomplete} is non-confluent.
In~\cite{overloading} we have identified these conditions as sufficient to ensure
completeness of the Canonical Normal Form procedure to test for equivalence among constraints.

\begin{theorem}
Let $P_p$ a complete and fully functional type class theory.
Then our CHR-based inference scheme infers principal types if the types arising are unambiguous.
\end{theorem}

%----------------------------------------------------------------------------------%

\mysection{Related Work and Conclusion} \label{sec:conc}

Simonet and Pottier~\cite{simonet-pottier-hmg} introduce HMG(X) a refined version of 
HM(X)~\cite{sulzmann-odersky-wehr:journal,sulzmann2000}
which includes among others type annotations. Their type inference approach
is based on the ``allow for more solutions'' philosophy.
Hence, they achieve complete type inference immediately. However, they only
consider tractable type inference for the specific case of equations as the only primitive constraints.

An approach in the same spirit is considered by
Hinze and Peyton-Jones~\cite{hinze+spj:derivable}. They sketch an extension of Haskell 
to allow for ``higher-order'' instances which logically correspond 
to nested equivalence relations.
As pointed out by Fax{\'e}n~\cite{haskell-pt}, in such an extended Haskell version 
it would be possible to type the program in Example~\ref{ex:no-pt}.
We believe this is an interesting avenue to pursue.
We are not aware of any formal results nor a concrete implementation of their proposal.

Pierce and Turner~\cite{345100} develop a local type inference scheme where
user-provided type information is propagated inwards to nodes which are below the annotation
in the abstract syntax tree.
Their motivation is to remove redundant annotations.
Note that Peyton-Jones and Shields~\cite{pratical-inference-rank-k} describe a particular instance of local type inference
based on the work by Odersky and L{\"a}ufer's\cite{237729}.
In our approach we are able to freely distribute type information across the
entire abstract syntax tree. Currently, we only distribute information about the types
of lambda-bound variables and type class constraints. 
We believe that our approach
can be extended to a system with rank-k types. We plan to pursue this
topic in future work.

In this paper, we have presented a novel inference scheme where the entire type inference problem is mapped to a set of CHRs.
Due to the constraint-based nature of our approach, we are able to make
available the results of inference for outer expressions while inferring the type of inner expressions.
We have fully implemented the improved CHR-based inference system as part of the
Chameleon system~\cite{chameleon}
%% which fully
%%incorporates the work described in this paper and will handle all the 
%%(typable) examples herein.
Our system improves over previous implementations such as Hugs and GHC.
For some cases, e.g.~unambiguous Haskell 98 programs, we can even state completeness.
We note that our improved inference scheme can host the type debugging techniques
described in~\cite{interactive,improved-debug}.

In future work, we plan to follow the path of Odersky and L{\"a}ufer~\cite{237729}
and compute (non-principal in general) solutions to subsumption problems.
We strongly believe that our improved CHR-based inference will be of high value for such an attempt.
Another alternative inference approach not mentioned so far is 
to only generate all necessary subsumption problems $\sigma_i \leq \sigma_a$
and wait for the ``proper'' moment to solve or check them. Of course, we still need to process them in a certain order
and might fail for the same reason we failed in Example~\ref{ex:p}.
Clearly, our constraint-based approach  allows us to ``exchange'' intermediate results among
two subsumption problems which may be crucial for successful inference.

\newpage

\appendix

\mysection{Variations}

Our exisiting translation to CHRs is slightly lazier than most inference
algorithms in the sense that we do not infer the types of let-bound variables
which are never called.

\begin{example}
Consider:

\begin{ttprog}
f x = \= let \=g = x x \\
      \> in  \>x
\end{ttprog}

We generate CHR rules that look like the following:
\bda{rcl}
f(t) & \simparrow & t = t_x \arrow t_x \\
g(t) & \simparrow & t_x = t_x \arrow t
\eda

Since the $f$ rule never calls $g$, the unsatisfiable constraint is not
introduced, and this program is considered well-typed.

\end{example}

This ``laziness'' can be problematic whenever we need to compare the inferred 
type of some function with its declared type.
Consider an annotated function $g$, nested within the definition of function
$f$, from which we generate an inference rule, 
$g(t,l) \simparrow g_a(t,l), C_i$, and an annotation rule, 
$g_a(t,l) \simparrow f(t',l')_{\ominus}, C_a$. 
It's possible that the types of some global variables are affected in $g$, but
not in $g_a$. 
In order for $g(t,l)$ and $g_a(t,l)$ to be equivalent, we depend on $g$'s
context, as called in the $g_a$ rule, to in turn call $g$ and introduce those
missing constraints.

\begin{example}
\label{ex:ann-ctxt}
Consider the following program.

\begin{ttprog}
f y = \=let \=g~::~Bool \\
      \>    \>g = y     \\
      \>in  \>'a'
\end{ttprog}

We generate the following (simplified) CHR rules:
\bda{rcl}
f(t,l) & \simparrow & t = Char,~l = (\llist{}, \llist{t_y}) \\
g(t,l) & \simparrow & t = t_y,~l = (\llist{t_y}, \llist{t_y}),~g_a(t,l) \\
g_a(t,l) & \simparrow & t = Bool,~l = (\llist{t_y}, \llist{t_y}),~f(t',l) 
\eda

Even though {\tt g}'s type annotation is acceptable, out subsumption check
would fail, because $g(t,l)$ and $g_a(t,l)$ are not equivalent wrt the $l$
component. 
Clearly, the $g$ rule implies $t_y = Bool$, but the $g_a$ rule does not.

\end{example}

We can remedy this situation by ensuring that all nested functions are called 
by their parent function. In this way, when we consider a function annotation,
the definition of the function becomes part of its context.

\begin{example}
We modify the program of Example~\ref{ex:ann-ctxt}, forcing {\tt f} to call
{\tt g}, but disregard its value.

\begin{ttprog}
f y = \=let \=g~::~Bool \\
      \>    \>g = y     \\
      \>in  \>const 'a' g
\end{ttprog}

The CHR rule associated with {\tt f} would now look something like:
\bda{rcl}
f(t,l) & \simparrow & t_{const} = a \arrow b \arrow a, a = Char, t = t_y \arrow a, \\
       &            & l = (\llist{}, \llist{t_y}),~g(b,l')
\eda

This solves our immediate problem, in that the constraints arising from
$g(t,l)$ and $g_a(t,l)$ are now equialent wrt $t$ and $l$.

Unfortunately this does not work in the case where the function we call is
recursive. Consider:

\begin{ttprog}
f y = \=let \=g~::~Bool \\
      \>    \>g = const y g \\
      \>in  \>'a' 
\end{ttprog}

Here, if {\tt f} were to call {\tt g}, the constraint generated to represent
{\tt g}'s type would be $g_a(t',l')$.
We then face the same problem, that from within $g_a$ we have no association
between $t_y$ and $Bool$.

\end{example}

Clearly, a syntactic transformation of the source program to introduce calls
to otherwise uncalled functions is not sufficient.
We must modify the CHR generation process to directly insert calls to the
inference constraints of functions which are not already called.

\begin{example}
We return to the rules generated in Example~\ref{ex:ann-ctxt}.
The following CHR rule is a modified version of the $f$ rule above which now
contains a call to $g$, further constraining the lambda-bound type variables.

\bda{rcl}
f(t,l) & \simparrow & t = Char,~l = (\llist{}, \llist{t_y}),~g(t',l)_{\ominus}
\eda

Using this modified rule, we see now that $g(t,l)$ and $g_a(t,l)$, are
equivalent, since $t_y = Bool$ in both.
The $\ominus$ mark on the $g$ constraint is not significant here, though it
does accomodates the simpler form of cycle breaking (by simply removing the
repeated constraint) than the equivalent unmarked constraint would.
\end{example}

\mysection{Monomorphic Recursive Functions} \label{sec:mon-rec-func}

\newcommand{\MRF}{\it MRF}
\newcommand{\ARF}{\it ARF}
\newcommand{\NRF}{\it NRF}

In case of monomorphic recursive functions (i.e.~recursive functions with no type annotation) 
we need to update our strategy for breaking cycles among CHRs (Definition~\ref{def:chr-cycle-removal}).
We denote by $\NRF$ the set of all non-recursive functions, by $\MRF$ the set of all recursive functions which carry no type annotations and
by $\ARF$ the set of all annotated recursive functions.

\begin{definition}[CHR Cycle Removal] 
Let $f(t,l)_{\ominus} \in C$ and $f(t',l')_{\ominus} \in C'$ where $f \in \NRF\cup\ARF$ and a derivation
            $... \rightarrowtail C \rightarrowtail ... \rightarrowtail C'$.
  Then, $... \rightarrowtail C \rightarrowtail ... \rightarrowtail C' \rightarrowtail_{CCR} C'-f(t',l')_{\ominus}$.

Let $f(t,l) \in C$ and $f(t',l')_{\ominus} \in C'$ where $f \in \MRF$ and a derivation
            $... \rightarrowtail C \rightarrowtail ... \rightarrowtail C'$.
  Then, $... \rightarrowtail C \rightarrowtail ... \rightarrowtail C' \rightarrowtail_{CCR} C'-f(t',l')_{\ominus},t'=t$.

Let $f(t,l) \in C$ and $f(t',l') \in C'$ where $f \in \MRF$ and $f(t',l')$ is
known to arise from the exact same program source location as $f(t,l)$, 
and given a derivation
            $... \rightarrowtail C \rightarrowtail ... \rightarrowtail C'$.
  Then, $... \rightarrowtail C \rightarrowtail ... \rightarrowtail C' \rightarrowtail_{Mono} C'-f(t',l'),t'=t$.

We assume that $\rightarrowtail_{CCR}$ and $\rightarrowtail_{Mono}$ are applied aggressively.
\end{definition}

Note that according to the definition of a $\rightarrowtail_{Mono}$ step, 
we should only ever break cycles amongst constraints representing unannotated, 
recursive function types when they arise from the same program location. 
Unifying the types of different calls to the same function is overly
restrictive, as the following example illustrates.

\begin{example}
Consider the following program.

\begin{code}
e = let f = f
    in  (f::Int, f::Bool)
\end{code}
\end{example}

The two {\tt f}s in the body of {\tt e} are calls to the same recursive,
unannotated function. i.e. ${\tt f} \in MRF$.
It would be unnecessarily restrictive, however, to aggressively apply 
$\rightarrowtail_{Mono}$ here and unify the types of the two {\tt f}s,
resulting in a type error. Indeed, these two {\tt f}s are not even part of the
same cycle.

We note that breaking cycles among constraints arising from the exact same program source location
is sufficient.

\begin{example}
Consider
\bp
f = ... f$_1$ ... f$_2$
\ep
where we added numbers $1$ and $2$ 
to different uses sites of {\tt f}.
Here is a sketch of type inference where we annotate $\rightarrowtail$ to refer to the number of
CHR steps applied so far.
$f \rightarrowtail^n \underline{f_1}, f_2 \rightarrowtail_f f_{1,1}, f_{1,2}, f_2$.
In the last step we reduce a call to $f$ at location $1$. Note that we make use of a refined
marking scheme. We keep track of the original
source location and add the location of the constraint introduced to the store to the existing
locations. We refer to~\cite{interactive} for the details of such a ``location-history'' aware refinement
of the CHR semantics.
Hence, applying rule \tlabel{Mono} twice will remove 
$f_{1,1}$ and $f_{1,2}$ (and equate the types of $f_{1,1}$ and $f_{1,2}$ with $f$). Similarly, we remove the cycles created by $f_2$.
\end{example}

We note that for  $\rightarrowtail_{CCR}$ 
there is no need to equate the $l$ component in case of $f\in \MRF$. The $l$ component is always set
at the use site of functions (see constraint generation rules \tlabel{Var-f}).
Equating the $l$ component would yield a strictly weaker system.

\begin{example}
Consider {\tt f x = ( f$_1$ ..., let g y = ... f$_2$ ...in g)} where we added numbers $1$ and $2$ 
to different (monomorphic) uses sites of {\tt f}. Here is a sketch of type inference
$f \rightarrowtail^n  ...f_1, ..., g ~\rightarrowtail^{n+m}  ... f_1, ..., f_2$ where natural
numbers $n$ and $m$ refer to the number of CHR steps applied so far.
Assume we fully equate $f_1$ and $f_2$. Note that their local $l_l$ components differ (because this component is always exact!)
Hence, type inference fails.
\end{example}

Note that (as before in case of $\ARF$ and $\NRF$) we only
break cycles among constraints which
carry the same marker. This yields a more precise method.
\begin{example}
Assume we have situation where
$$
... \rightarrowtail f(Int), ... \rightarrowtail ... \rightarrowtail f(t)_{\ominus}, ...
$$
Removing $f(t)_{\ominus}$ and adding $t=Int$ might be too restrictive.
\end{example}

Note that by construction $\rightarrowtail_{Mono}$ never applies to $\ARF$ and $\NRF$.
Any potential cycle will be eventually broken.
We can re-establish Lemmas~\ref{le:cchr-termination} and and state a slightly
stronger Lemma~\ref{le:rmc-soundness} which guarantee
soundness of our CHR-based inference approach.

\begin{lemma}[CCR-Mono Soundness] \label{le:rmc-soundness2}
Let $P$ be a set of CHRs and $C$ and $C'$ two constraints such that $C\rightarrowtail^*_P C'$.
Then $P \models \bar{\exists}_{\stv(C)} C' \implies C$.
\end{lemma}

Note that $P \models C \implies \bar{\exists}_{\stv(C)} C'$ does not hold anymore.
We may reject typable programs because we stricly enforce the \tlabel{Mono} rule.

\begin{example}
Consider
\bp
h x = (h$_1$ 'a') \&\& (h$_2$ True)
\ep
Here is a sketch of type inference.
\bda{ll}
 & h(t) \\
\rightarrowtail & t=tx\arrow Bool, h_1(Char\arrow Bool), h_2(Bool\arrow Bool) \\
\rightarrowtail & t=tx\arrow Bool, 
     Char\arrow Bool=tx' \arrow Bool, h_{1,1}(Char \arrow Bool), \\
   & h_{1,2}(Bool\arrow Bool),
     h_2(Bool\arrow Bool) \\
\rightarrowtail_{Mono} & 
t=tx\arrow Bool, 
     Char\arrow Bool=tx' \arrow Bool, Char \arrow Bool=Char \arrow Bool, \\
    & h_{1,2}(Bool\arrow Bool),
     h_2(Bool \arrow Bool) \\
\rightarrowtail_{Mono} & 
t=tx\arrow Bool, 
     Char\arrow Bool=tx' \arrow Bool, Char \arrow Bool=Char \arrow Bool, \\
    & Char \arrow Bool=Bool\arrow Bool,
     h_2(Bool \arrow Bool) \\
\leftrightarrow & False
\eda
\end{example}

It is interesting to note is that
our type inference scheme for recursive function is more 
relaxed compared to 
the one found in some other established type checkers.

\begin{example} \label{ex:polyrec2}
Consider the following program. 
\begin{code}
e :: Bool
e = g 
f :: Bool -> a
f = g 
g = f e
\end{code}
In the case of GHC, the following error reported is:
\begin{ttprog}
mono\=-rec.hs:5: \\
    \>Coul\=dn't match `Bool -> a' against `Bool'   \\
    \>    \>Expected type\=:~Bool -> a  \\
    \>    \>Inferred type\>:~Bool   \\
    \>In the definition of `f':~f = g
\end{ttprog}

The problem reported here stems from the fact that within the mutually
recursive binding group consisting of {\tt e, f} and {\tt g}, {\tt f} is
assigned two ununifiable types, $Bool$ and $Bool \arrow a$: the first because it
must have the same type as {\tt g}, which according to {\tt e} must be $Bool$;
and the second because of its type declaration.

Our translation scheme is more liberal than this, in that {\tt g}'s type within
{\tt e} and {\tt f} may be different.
Essentially, we only require that the type of a variable be identical at all
locations within the mutually recursive subgroup if a type declaration has
been provided for that variable.

(Simplified) Translation of the above program to CHRs yields.
$$
\begin{array}{lcl}
e_a(t) & \simparrow & t = Bool \\
e(t)   & \simparrow & g(t) \\
f_a(t) & \simparrow & t = Bool \arrow a \\
f(t)   & \simparrow & g(t) \\
g(t)   & \simparrow & e_a(t_e), f_a(t_f), t_f = t_e \arrow Bool
\end{array}
$$
It is clear from the above that there are no cycles present amongst these
rules. 
We can use them to successfully infer a type for any of the variables in the 
program.

In fact, our handling of binding groups is similar to~\cite{jones99typing} where
type inference of binding groups proceeds as follows:
\begin{enumerate}
 \item  Extend the type environment with the type signatures.
        In this case {\tt f::forall a. Bool -> a} and {\tt e::Bool}.
 \item Do type inference on the bindings without type signatures,
        in this case {\tt g = f e}.
    Do generalisation too, and extend the environment, giving
        {\tt g~::~forall a. a}.
 \item Now, and only now, do type inference on the bindings with
        signatures.
\end{enumerate}
\end{example}

\end{document}